\documentstyle [prl,aps,floats,twocolumn]{revtex}
\input psfig.tex
\begin{document}

\def\ba{\begin{eqnarray}}
\def\ea{\end{eqnarray}}
\def\be{\begin{equation}}
\def\ee{\end{equation}}
\def\({\left(}
\def\){\right)}
\def\[{\left[}
\def\]{\right]}
\def\lagrange {{\cal L}}
\def\del {\nabla}
\def\d {\partial}
\def\Tr{{\rm Tr}}
\def\half{{1\over 2}}
\def\fourth{{1\over 8}}
\def\bibi{\bibitem}
\def\S{{\cal S}}
\def\xx{\mbox{\boldmath $x$}}
\newcommand{\labeq}[1] {\label{eq:#1}}
\newcommand{\eqn}[1] {(\ref{eq:#1})}
\newcommand{\labfig}[1] {\label{fig:#1}}
\newcommand{\fig}[1] {\ref{fig:#1}}
\newcommand\bigdot[1] {\stackrel{\mbox{{\huge .}}}{#1}}
\newcommand\bigddot[1] {\stackrel{\mbox{{\huge ..}}}{#1}}

\twocolumn[\hsize\textwidth\columnwidth\hsize\csname @twocolumnfalse\endcsname
\title{Classification of Normal Modes for Multiskyrmions} \author{Kim
Baskerville and Robin
Michaels} \address{Department
of Mathematical Sciences, Durham University, Science Laboratories,
South Road, Durham DH1 3LE}
\date\today 
\maketitle

\begin{abstract}
  The normal mode spectra of multiskyrmions play a key role in their
  quantisation. We present a general method capable of predicting all
  the low-lying vibrational modes of known minimal energy
  multiskyrmions. In particular, we explain the origin of the higher
  multipole breathing modes, previously observed but not understood.
  We show how these modes may be classified according to the symmetry
  group of the static solution. Our results provide strong hints that
  the $N$-skyrmion moduli space, for $N>3$, may have a richer
  structure than previously thought, incorporating $8N-4$ degrees of
  freedom.
\end{abstract}
\vskip .2in
]

\section{Introduction}

The Skyrme model of nuclear physics \cite{Skyrme} describes baryons as
topological solitons in a non-linear field theory of $\pi$-mesons. It
has enjoyed qualitative success in describing both single nucleon
properties and the nucleon-nucleon interaction
\cite{ANW,JackRho,Jacks}. It is thus of interest to try to apply the
model to larger nuclei. In theory, it should be possible to calculate
the binding energies and gamma ray spectra of all nuclei, without
needing to introduce additional free parameters. This is a very
attractive prospect. However, solutions of the classical field theory
must be quantised before any comparison to the real world can be made.

The Skyrme model is not renormalisable, so there is little hope of a
full treatment as a quantum field theory. Instead, a semiclassical
quantisation is usually attempted, based on a limited number of
degrees of freedom about a given classical solution. The earliest
efforts at quantisation included only zero mode collective
coordinates: spin and isospin. For a single skyrmion, this produces a
description of nucleons and the $\Delta$ resonance in modest agreement
with experiment \cite{AN,ANW}. However, for solutions of higher baryon
number $B$, a simple collective coordinate quantisation includes
effects of order $\hbar^{2}$, while ignoring effects of order $\hbar$.
For this reason, there has recently been a growing acceptance that it
is necessary to include at least some low-lying vibrational modes.
This requires a better understanding of the structure and dynamics of
multi-skyrmion configurations, at least in the immediate neighbourhood
of the minimal energy solutions.

There has been considerable recent progress in this direction, based
on Manton's old idea of representing low energy solitonic excitations
as motion on a finite dimensional moduli space. Studies by Leese et al
\cite{Leese} and Walet \cite{Walet}, employing an instanton
approximation for the Skyrme fields, gave encouraging results.
Recently, Barnes et al directly computed the normal mode spectra of
$B=2$, 3 and 4 multiskyrmions \cite{deut,alpha,conf}. A remarkable
structure emerged in all three spectra. The lowest frequencies
corresponded to known attractive channel scatterings; the next mode up
was the `breather' (a trivial size fluctuation), followed by various
higher multipole breathing modes.

Barnes et al were also able to classify the vibrations according to
the symmetry groups of the respective static solitons. Remarkably, the
$B=4$ vibrational modes below the breather fell into representations
exactly corresponding to those for small zero-mode deformations of the
BPS 4-monopole solution. The same phenomen occurred for the low
frequency vibrations of the deuteron. This led to further
investigation of the connection between BPS monopoles and Skyrmions,
which has now been understood by Houghton et al in terms of rational
maps \cite{Houghton}. In fact, Houghton et al were able to show that
the correspondence should continue for any baryon number (monopole
charge), and were thus able to predict the lowest $4B-7$ vibrational
modes for $B=3$ and 7. Their $B=3$ predictions were then confirmed by
\cite{conf}.

The higher multipole breathing modes were not so well understood. In
the current Letter, we shall present a simple geometrical explanation
for these modes, which relies only on the symmetry of the static
solutions \cite{Battye}. We predict $4B-7$ such modes for a
multiskyrmion of baryon number $B$, and show how these may be
classified as representations of the symmetry group of the static
soliton. Our predictions match exactly all known vibrations for $B=2$,
3 and 4; plus we predict an additional triplet of modes for $B=4$. We
also make detailed predictions for $B=5$, 6 and 7.

Our idea is extremely simple, but has far-ranging consequences.
Perhaps the most startling of these is the implication that the
$N$-skyrmion moduli space may be of a higher dimension than was
previously thought. Interestingly, our theory is in agreement with the
$B=3$ results of \cite{conf}, which already contradicted the
``standard wisdom'' in this regard. We are thus able to clarify an
outstanding puzzle. Overall, when added to the already considerable
progress made by \cite{deut,alpha,conf,Houghton,Battye}, this
represents a major new insight into the moduli space approach.

\section{Classification of Multipole Breathing Modes}

Classical multiskyrmion solutions are now known up to baryon number
$B=9$ \cite{Braaten,Battye}. All display considerable symmetry: the
$B=1$ solution is spherical, while the deuteron has axial symmetry.
For $B \ge 3$, the minimal energy multiskyrmions are polyhedra, with
$2B-2$ faces. For example, $B=3$ is a tetrahedron and $B=4$ a cube.
Baryon density is peaked at the vertices of these polyhedra, and to a
lesser extent along the edges joining them. Lines of zero baryon
density run out from the origin through the midpoints of each face.
These lines are rather special in another way. Consider the inverse
map from the field space $SU(2)$ back to real space. In general, for a
configuration of baryon number $B$, each point in field space maps to
$B$ points in real space. But there are certain special field values
which map to fewer points in real space. For a minimal energy
multiskyrmion, the image of these points, which we shall refer to as
the branch locus, corresponds to the zero baryon density lines (or
branch lines) running through each face. Since there are $2B-2$ faces,
there are also $2B-2$ branch lines. Note that for this purpose, a line
of zero baryon density which runs straight through the origin counts
as two branch lines.

Our idea is to consider vibrations of the branch locus, while leaving
the baryon density distribution of the multiskyrmion roughly intact.
We leave the branch locus fixed at the origin, but allow the locations
at which the branch lines intersect a sphere of fixed radius to
vary. There are $2B-2$ branch lines, each of which has two angular
degrees of freedom, giving a total of $4B-4$ modes. Three of these
will correspond to global rotation, so we are left with $4B-7$
non-trivial vibrations.  The latter clearly correspond to complex
breathing motions: two branch lines moving towards each other will
compress the baryon density between them, whereas motion away from
baryon density peaks will cause expansion.

The non-trivial vibrational modes must lie in multiplets (of
degenerate frequency), transforming under irreducible representations
of the symmetry group of the static soliton. These representations can
be found by decomposing the $(4B-4)$-dimensional representation formed
by the angles of the branch lines in spherical polar coordinates. The
irreducible component(s) corresponding to the rotational zero modes
can then be removed, leaving only the true vibrational modes. Note
that while the spherical polar angles form the most convenient general
definition of the $(4B-4)$-dimensional representation, in practice it
is usually possible to find a simpler parametrisation for the movement
of the branch lines. This is the case for most examples we will
consider here.

Let us now consider detailed predictions for individual
multiskyrmions, beginning with the deuteron. The minimal energy $B=2$
solution is a torus: with axial symmetry, plus a reflection symmetry
in the plane of the torus. The symmetry group is $D_{\infty h}$, axial
symmetry extended by inversion. The two branch lines occupy the axis
of symmetry; take this to be the $z$-axis. Then the $x$- and $y$-axes
at some points $z = \pm \alpha$ are equivalent to the 4-dimensionsal
anglular basis. A simple computation reveals that the characters of
this representation are uniformly zero, other than the identity with
value 4, and rotations by angle $\theta$ with value $4
\cos \theta$. So using the notation of \cite{deut}, we have two
2-dimensional irreducible representations: $1^{+}$ and $1^{-}$. The
former corresponds to rotations around axes perpendicular to $z$, so
we are left with $1^{-}$ as a true vibration.  This is exactly the
mode found by \cite{deut}. It is a ``dipole'' breathing motion, where
one side of the torus inflates while the other is compressed. The
two-fold degeneracy corresponds to motion aligned along the $x$- or
$y$-axes.

The classical $B=3$ multiskyrmion is a tetrahedron; its branch lines
pass through a (dual) tetrahedron. There is no obviously simple way to
parametrise the 8-dimensional angular representation in this case
(although we have performed this calculation and checked the results
below).  However, the possible motions of the branch lines are quite
limited, and are easy to see. One can have a dipole motion, where
three of the branch lines move towards the fourth. There are four
obvious directions for this motion, but exciting all four at once
gives a trivial size fluctuation, so only three are independent. A
quick check of the symmetry of this vibration under the tetrahedral
group $T_{d}$ shows its representation to be $F_{2}$, using the
notation of Hamermesh \cite{Hamermesh}. (We will use this notation
throughout, except where otherwise explicitly noted.) It is also
possible to split the branch locus into two pairs of lines, with the
lines in each pair moving towards each other. This creates a
``quadrupole'' breathing motion, where two opposite edges of the
multiskyrmion are compressed, while the other four inflate; and vice
versa. This motion has an obvious threefold basis, but only two
directions are independent, so the corresponding representation is
$E$. Again, these two modes correspond to those observed by Barnes et
al \cite{conf}.

The $B=4$ multiskyrmion is a cube; symmetry group $O_{h}$. The branch
lines are just the Cartesian axes. We label the irreducible
representations by those of the group of rotations of a cube ($O$),
with superscripts indicating parity. Decomposition of the
12-dimensional branch line angle representation gives $F_{1}^{+}$,
$F_{1}^{-}$, $F_{2}^{+}$ and $F_{2}^{-}$, all 3-dimensional. The
$F_{1}^{+}$ corresponds to rotational zero-modes. $F_{1}^{-}$ is a
dipole breathing motion, and $F_{2}^{+}$ a quadrupole, similar to the
modes described above for $B=2$ and $B=3$. Both of these modes were
observed, with the right symmetries, by \cite{alpha}. The remaining
vibration, $F_{2}^{-}$, corresponds to a twisting motion: grab
diagonally opposite corners of the cube, and twist them in opposite
directions. This mode was not seen by \cite{alpha}, but it may well
have considerably higher energy than the other two. Finding it would
provide striking confirmation of our theory.

\begin{figure}[htbp]
  \begin{center} 
   \leavevmode 
\psfig{file=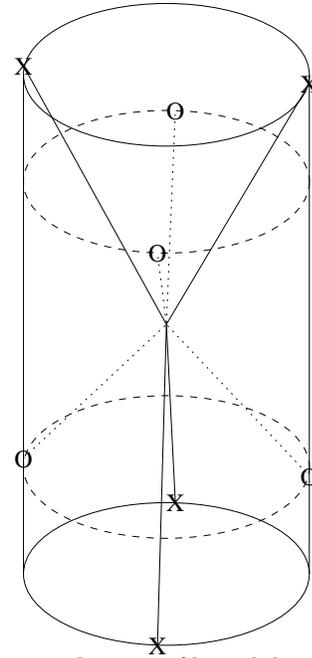,width=1.6in}
    \caption{Schematic diagram of branch lines of $B=5$
    multiskyrmion. Solid lines pass through square faces; dotted lines
    through pentagons.}
  \label{fig:d2d} 
  \end{center}
\end{figure}

The next two multiskyrmions have rather less symmetry: $D_{2d}$ and
$D_{4d}$ respectively. Pictures of them may be found in \cite{Battye}.
The $B=5$ solution has four square and four pentagonal faces. The
branch lines form two slightly distorted tetrahedral configurations:
see Figure~\ref{fig:d2d} for a schematic diagram. The symmetry group
$D_{2d}$ maps each of these ``tetrahedra'' to themselves; they cannot
be interchanged. The 16-dimensional angular representation can
therefore be reduced to two identical 8-dimensional
representations. Each of these can then be decomposed into $A_{1}$,
$A_{2}$, $B_{1}$, $B_{2}$ and $E$ (twice). The rotational zero modes
are $E$ and $A_{2}$. Again, we have dipole and quadrupole type
breathing motions.  Since the $B=5$ multiskyrmion is slightly
elongated about one axis, both the dipole and quadrupole motions split
into a singlet and a doublet: $B_{2}$ and $E$ for the dipole, $A_{1}$
and $E$ for the quadrupole. There are three 1-dimensional twisting
motions, two $B_{1}$'s and an $A_{2}$. The remaining motions
correspond to two groups of four axes each vibrating in a dipole or
quadrupole, but with the two groups out of phase. Hence the remaining
$B_{2}$ is an axial dipole anti-dipole, $E$ a transverse dipole
anti-dipole, and $A_{1}$ an axial quadrupole anti-quadrupole.

The $B=6$ multiskyrmion has two square faces and eight pentagonal
faces. Arrange four pentagons around each square, then rotate one of
the squares by $45^{\circ}$ so that the two halves can be fitted
together. This configuration, like $B=5$, is slightly elongated in one
direction. The symmetry group $D_{4d}$ is not included in Hamermesh
\cite{Hamermesh}, so we write out its character table (see
Table~\ref{tab:char}) in order to define notation for the
representations. Our standard decomposition of the 20-dimensional
representation gives $1^{+}$, $1^{-}$, $A^{+}$,
$A^{-}$, $C$ (twice), $B^{+}$ (three times) and $B^{-}$ (three times). The rotational modes to be discarded are
$A^{+}$ and $B^{-}$. The situation is now too
complicated for us to be confident of identifying the kind of motion
corresponding to each representation. However, we can make a few
obvious assignments. There will be axial and transverse dipole motions
($1^{-}$ and $A^{-}$); also axial and transverse quadrupole
modes ($1^{+}$ and $C$). $B^{+}$ and the remaining
$C$ could be twisting modes. Bending or wobbling the $z$-axis
(the branch lines passing through the two squares) while leaving the
other axes fixed should give $B^{+}$ and $B^{-}$
respectively. This leaves two modes unidentified; it will be
interesting to examine them when they are computed.
\begin{table}[htbp]
  \begin{center}
    \leavevmode
    \begin{tabular}[c]{|c|c|c|c|c|c|c|c|} \hline
\ & $E$ & $C_{4}$ & $C_{2}$ & $S_{8}$ & $S_{8}^3$ & $\sigma C$ &
$\sigma S$ \\ \hline
$1^{+}$ & 1 & 1 & 1 & 1 & 1 & 1 & 1 \\
$1^{-}$ & 1 & 1 & 1 & -1 & -1 & 1 & -1 \\
$A^{+}$ & 1 & 1 & 1 & 1 & 1 & -1 & -1 \\
$A^{-}$ & 1 & 1 & 1 & -1 & -1 & -1 & 1 \\
$B^{+}$ & 2 & 0 & -2 & $\sqrt{2}$ & $-\sqrt{2}$ & 0 & 0 \\
$B^{-}$ & 2 & 0 & -2 & $-\sqrt{2}$ & $\sqrt{2}$ & 0 & 0 \\
$C$ & 2 & -2 & 2 & 0 & 0 & 0 & 0 \\ \hline
    \end{tabular}
    \caption{Character table for $D_{4d}$ }
    \label{tab:char}
  \end{center}
\end{table}

Finally, we consider $B=7$. This is a perfect dodecahedron, with
symmetry group $I_{h}$. We label the representations of this group by
those of ${\rm Alt}_{5}$, the even permutations of five objects, with
superscripts to indicate parity. We make no attempt to identify the
physical appearances of these modes. We can, however, predict their
symmetries. Our decomposition gives $5^{+}$, $5^{-}$, $4^{+}$,
$4^{-}$, $3^{+}$ and $3^{-}$. The triplet $3^{+}$ corresponds to the
discarded zero-modes. Interestingly, these representations correspond
to those predicted by \cite{Houghton} for the $4B-7$ lower frequency
scattering modes, although these authors do not give parity
assignments. This also happens for the $B=3$ tetrahedron, which we
attribute to the self-duality of a tetrahedron. In all other cases,
however, the $4B-7$ higher breathing modes have quite different
symmetries to the $4B-7$ scattering modes.

\section{Discussion and conclusions}

We have proposed a simple explanation of the origin of the higher
multipole breathing modes observed in multiskyrmions; namely, that
they correspond to vibrations of the branch locus, or lines of zero
baryon density. Our results are summarised in Table~\ref{tab:predict}.
We predict $4B-7$ such modes for a multiskyrmion of baryon number $B$
($4B-6$ for the deuteron).  Together with the $4B-7$ lower frequency
scattering modes, plus nine zero modes and one trivial breather, this
gives a total of $8B-4$ modes ($8B-3$ for the deuteron). This would
appear to resolve a long-standing ``counting problem''.

A single skyrmion has 6 degrees of freedom, which leads to the naive
expectation that an $N$-nucleon system should have $6N$ degrees of
freedom, and hence be described by a $6N$-dimensional moduli space. It
can be argued that this dimensionality should be increased by one,
since all Skyrme configurations have a trivial size fluctuation.
Since minimal energy solutions for $B>1$ are single large
solitons, they therefore have a maximum of 9 zero modes. This gave
rise to the hope that multiskyrmions would have exactly $6B-9$
low-lying vibrational modes; that these vibrations might in fact
correspond directly to the ``broken zero modes'' of well-separated
skyrmions. The results of Barnes et al for $B=2$ and $B=4$ seemed to
support this notion; they found just exactly the right number of
vibrations in each case. However, they found one too many mode for
$B=3$, and in a multiplet which prevented separating this ``extra mode''
from the others. Since the lower half of this spectrum perfectly
matched the predictions of \cite{Houghton} however, it was hard to
discard their results. We now predict exactly the multiplets Barnes et
al observed. This is strong evidence against a $(6B+1)$-dimensional
moduli space. 

So what is going on? It would seem either that the moduli space
approach is wrong, or that Skyrme configurations (for $B \ge 3$) have
$2B-5$ more degrees of freedom than was previously thought. This means
that knowing the position and orientations of $N$ skyrmions is not
sufficient information to determine the field everywhere in space.
What other structure could there be? One possibility is that the
branch locus contains additional information. Another is that additional
dynamical variables are required, for example arising from
interactions between angular velocities of different skyrmions.
Further speculation is beyond the scope of this Letter, but the
current results certainly indicate that further investigation of the
structure of the branch loci would be worthwhile.

RM would like to thank Wojtek Zakrzewski for useful discussions
on the background to this work. KB is supported by PPARC; RM partially
by the Department of Mathematical Sciences, Durham University.

\begin{table}[htbp]
  \begin{center}
    \leavevmode
    \begin{tabular}{|c|c|c|c|c|} \hline
B & Symmetry & Mode & Degeneracy & Description \\ \hline
2 & $D_{\infty h}$ & $1^{-}$ & 2 & dipole \\ \hline
3 & $T_{d}$ & $F_{2}$ & 3 & dipole \\
\ & \ & $E$ & 2 & quadrupole \\ \hline
4 & $O_{h}$ & $F_{1}^{-}$ & 3 & dipole \\
\ & \ & $F_{2}^{+}$ & 3 & quadrupole \\ 
\ & \ & $F_{2}^{-}$ & 3 & twist \\ \hline
5 & $D_{2d}$ & $B_{2}$ & 1 & axial dipole \\
\ & \ & $E$ & 2 & transverse dipole \\ 
\ & \ & $A_{1}$ & 1 & axial quadrupole \\ 
\ & \ & $E$ & 2 & transverse quadrupole \\ 
\ & \ & $B_{1}$ & 1 & twist \\ 
\ & \ & $B_{1}$ & 1 & twist \\ 
\ & \ & $A_{2}$ & 1 & twist \\ 
\ & \ & $B_{2}$ & 1 & axial `anti-dipole' \\ 
\ & \ & $E$ & 2 & transverse `anti-dipole' \\ 
\ & \ & $A_{1}$ & 1 & axial `anti-quadrupole' \\ \hline
6 & $D_{4d}$ & $1^{-}$ & 1 & axial dipole \\
\ & \ & $B^{+}$ & 2 & transverse dipole \\
\ & \ & $1^{+}$ & 1 & axial quadrupole \\
\ & \ & $C$ & 2 & transverse quadrupole \\
\ & \ & $A^{+}$ & 1 & twist \\
\ & \ & $C$ & 2 & twist \\
\ & \ & $B^{+}$ & 2 & $z$-axis bend \\
\ & \ & $B^{-}$ & 2 & $z$-axis wobble \\
\ & \ & $B^{+}$ & 2 & \ \\
\ & \ & $B^{-}$ & 2 & \ \\ \hline
7 & $I_{h}$ & $5^{-}$ & 5 & dipole \\
\ & \  & $5{+}$ & 5 & quadrupole \\
\ & \  & $4^{+}$ & 4 & \ \\
\ & \  & $4^{-}$ & 4 & \ \\
\ & \  & $3^{-}$ & 3 & \ \\ \hline
    \end{tabular}
    \caption{Summary of Predictions for $B=2$ to $B=7$}
    \label{tab:predict}
  \end{center}
\end{table}

\end{document}